# Oxygen Nuclear Magnetic Resonance Study of $RuSr_2EuCu_2O_8$

S. Krämer[1,†] and G. V. M. Williams[1,2]

[1]*2. Physikalisches Institut, Universität Stuttgart, D-70550 Stuttgart, Germany.*
[2]*Industrial Research Limited, P.O. Box 31310, Lower Hutt, New Zealand.*

## ABSTRACT

We report the results from a $^{17}O$ Nuclear Magnetic Resonance (NMR) study of the high temperature superconducting cuprate (HTSC), $RuSr_2EuCu_2O_8$. This compound has recently been found to exhibit the coexistence of magnetic order and superconductivity. At 9 T, the magnetic order in the $RuO_2$ layers is predominately ferromagnetic. We measured the $^{17}O$ NMR spectra from 10 K to 330 K and find the occurrence of an increasing $^{17}O$ NMR shift with increasing temperature for temperatures above ~200 K, providing direct evidence that $RuSr_2RCu_2O_8$ is similar to a very underdoped HTSC. The analysis of the NMR shift and spectral width of the $^{17}O$ NMR line gives a small but measurable hyperfine field of ~9 mT from the Ru moment at the oxygen site in the copper oxygen planes. We show that the $^{17}O$ NMR linewidth is governed by dipolar contributions from the Ru moment.

† Corresponding author
  Phone: +49 (0)711 685-5177
  Fax:   +49 (0)711 685-5285
  e-mail: `s.kraemer@physik.uni-stuttgart.de`

**INTRODUCTION**

There have been a number of reports of the coexistence of magnetic order and superconductivity in the ruthenate-cuprates, $RuSr_2R_{2-x}Ce_xCu_2O_{10+\delta}$ and $RuSr_2RCu_2O_8$ where R is Gd, Eu or Y [1-19]. These compounds were first synthesized by Bauernfeind, Widder and Braun [13,14]. They are unique in that, unlike previous compounds, the magnetic order occurs at temperatures much greater than the superconducting transition temperature, $T_c$. Recent reports have focused on $RuSr_2RCu_2O_8$ [3-13,16] where a bulk Meissner phase is found to exist for temperatures below ~35 K for R=Gd and below ~10 K for R=Eu while the magnetic ordering transition occurs at a much higher temperature of ~132 K. The unit cell of $RuSr_2RCu_2O_8$ is similar to the high temperature superconducting cuprate (HTSC), $YBa_2Cu_3O_{7-\delta}$, in that it contains two $CuO_2$ layers but the CuO chains are replaced by a $RuO_2$ layer. However, there is an additional complication in that the crystallite domain size is small and there are coherent rotations of the $RuO_6$ octahedra about the c-axis leading to a superstructure and subdomains of 5 nm to 20 nm in diameter [5,8]. It has been suggested on the basis of transport, heat capacity and microwave measurements that the electronic behavior of $RuSr_2RCu_2O_8$ is similar to an underdoped HTSC [6,18] However, nuclear magnetic resonance (NMR) measurements have been interpreted in terms of $RuSr_2RCu_2O_8$ being comparable to an optimally doped HTSC [24].

Some reports have concluded that the magnetic order in $RuSr_2RCu_2O_8$ is ferromagnetic in the $RuO_2$ layers [3-6,9,10,13] in which case there should be a competition between the superconducting and magnetic order parameters possibly resulting in a spontaneous vortex phase or a spatial modulation of the respective order parameters [7]. However, powder neutron diffraction studies on $RuSr_2GdCu_2O_8$ and $RuSr_2YCu_2O_8$ showed that the low-field magnetic order is predominately antiferromagnetic (~1 $\mu_B$/Ru) and there is a spin-flop transition that gradually occurs for magnetic fields greater than 0.4 T [11,12]. It was found that the antiferromagnetic order arises from Ru in the $RuO_2$ layers and there is no evidence of long range magnetic order in the $CuO_2$ layers. We have shown that the magnetization data from measurements on $RuSr_2EuCu_2O_8$ are consistent with the low-field magnetic order being predominately antiferromagnetic and the high-field magnetic order being predominately ferromagnetic [16]. The weak ferromagnetic component at zero field is estimated to be ~0.05 $\mu_B$/Ru [16].

One of the important unanswered questions is the extent of exchange coupling from the moments in the magnetic $RuO_2$ layers to the superconducting $CuO_2$ layers. In particular it is not known if the superconductivity is suppressed in the $CuO_2$ layers due to pairbreaking via the Ru moments. It has been suggested from a magneto-transport measurement on $RuSr_2GdCu_2O_8$ above the magnetic ordering temperature that there is strong exchange coupling from the Ru moments to the $CuO_2$ planes, which should significantly suppress superconductivity [9]. However, it was also stated that the data could be interpreted in terms of itinerant ferromagnetism and significant electronic conduction in the $RuO_2$ layers. This second scenario is possible, as recent band structure calculations have indicated that the conduction band is metallic but the calculations indicate that the magnetism could arise via a double exchange mechanism [20]. Some indication of exchange coupling was provided by a Gd electron spin resonance (ESR) of RuSr$_2$GdCu$_2$O$_8$ where the exchange field at the Gd site was found to be ~60 mT [10]. This is less than the reported dipolar field of 70 mT to 100 mT obtained by muon spin rotation (μSR) measurements on RuSr$_2$GdCu$_2$O$_8$ [4]. A direct measurement of exchange coupling between the CuO$_2$ layers and the Ru moment is also provided by Cu NMR [21,22]. The results of a recent Cu NMR study would appear to suggest that there is only a weak contribution of the Ru moment to the total hyperfine shift at the planar Cu site in this material. In this particular NMR study an upper limit of 200 mT has been estimated for any internal field contribution to the NMR lineshift at this site.

Recent reports of $^{99,101}$Ru zero-field NMR and of X-ray absorption near edge spectroscopy (XANES) [21,22,23] have shown that the valence state of Ru in RuSr$_2$RCu$_2$O$_8$ (R=Gd,Y) is a mixed valence of Ru$^{4+}$ and Ru$^{5+}$ with fractions of 40-50 % Ru$^{4+}$ and 60-50 % Ru$^{5+}$. Since the doping mechanism of this material is an intra unit-cell charge transfer of holes from the CuO$_2$ plane to the RuO$_2$ plane without any change in stochiometry, the electronic and magnetic properties of this material are strongly correlated. In particular, there should be a correlation between the hole concentration on the CuO$_2$ planes and the valence state of Ru. In this context, another important question is, therefore, what is the hole concentration of the CuO$_2$ planes? There has been a recent Cu NMR spin-lattice relaxation report which suggested that the hole concentration on the CuO$_2$ planes in RuSr$_2$YCu$_2$O$_8$ (T$_c$= 23 K) is close to that of an optimally doped HTSC with a hole concentration, p, of ~0.16 [24]. However, thermopower measurements on RuSr$_2$GdCu$_2$O$_8$ (T$_c$=45 K) and RuSr$_2$EuCu$_2$O$_8$ (T$_c$=32 K) have been interpreted in terms of these materials being comparable to an underdoped HTSC (p~0.07) [6,18]. It should be noted that the comparison with HTSC thermopower data is indirect and qualitative. This can be compared with the analysis of structural data where it has been suggested that RuSr$_2$GdCu$_2$O$_8$ should be very overdoped [5]. Furthermore, the large increase in T$_c$ with decreasing rare earth ionic radii is unusual and opposite to that observed in RBa$_2$Cu$_3$O$_7$, where T$_c$ decreases by ~3 K when the ionic radii is decreased from Gd to Y [25].

In this paper we report the results from a $^{17}$O NMR study on RuSr$_2$EuCu$_2$O$_8$ where there is a coexistence of superconducting and magnetic order. NMR is an ideal technique to probe the ruthenate-cuprates because it is possible to probe both the spin and charge dynamics without the problems of impurity phases experienced by other techniques. In particular, it has been found that the temperature dependence of the $^{17}$O Knight shift is a precise indicator of the doping state in the HTSC. The current measurements were performed on RuSr$_2$EuCu$_2$O$_8$ rather than the more-studied RuSr$_2$GdCu$_2$O$_8$ because the moment from Gd is known to lead to a significant broadening of the NMR spectra as well as affecting the spin-lattice relaxation rates. Unlike the other HTSC, it was not possible to c-axis align RuSr$_2$RCu$_2$O$_8$ and hence the measurements were performed on an unaligned powder sample. We show below that $^{17}$O NMR spectra are observed above and below the magnetic ordering temperature.

**EXPERIMENTAL DETAILS**

The sample was prepared in exactly the same manner as $RuSr_2GdCu_2O_8$ [3,13,16]. The sample was $^{17}O$ exchanged by annealing the sample in 50% enriched $^{17}O$ gas at 700 °C for six hours. The $^{17}O$ NMR measurements were made at an external field of 9 T using a home built pulsed NMR spectrometer with a continuous flow cryostat [26]. The spectra were obtained by summing up the Fourier transformed spectra of the second half of the echo at different carrier frequencies. $^{17}O$ NMR spin-lattice relaxation measurements were made using an inversion-recovery pulse sequence. Since we used an unoriented powder sample, we expect a powder pattern for the satellite transitions of the $^{17}O$ nuclear spin 5/2 system and a well defined central transition which is predominately excited by the rf-field. Therefore, the $^{17}O$ NMR relaxation data were fitted according to the relaxation after partial inversion of the central transition of a spin 5/2 system [27],

$$M(t) = M_0\left(1 - \Gamma\left(\frac{1}{35}\exp(-2W_1t/3) + \frac{8}{45}\exp(-4W_1t) + \frac{50}{63}\exp(-10W_1t)\right)\right), \quad (1)$$

using the convention of $W_1 = {}^{17}T_1^{-1}$ introduced by Martindale *et al.* [28]. $\Gamma$ denotes the inversion ratio, $\tau$ is the time between the inversion pulse and the detection sequence and $M_0$ gives the equilibrium magnetization. The Cu nuclear quadrupole resonance (NQR) spectrum was measured at room temperature to enable a comparison of the Cu NQR frequency with that in the HTSC. The Cu NQR spectra were measured in the same way as the $^{17}O$ spectra and using a µ-metal shielded set-up in order to minimize the influence of external magnetic fields.

The ac and dc magnetization measurements were made using a SQUID magnetometer and, in the case of the ac measurements, a dc magnetic field of zero T (note that due to the remnant magnetization of the superconducting magnet, the field with the applied current off is ±0.6 mT), an ac magnetic field of 5 µT and a frequency of 1 kHz. The low-field dc magnetization was measured at 4 mT to ensure that there were no sample container effects or effects due to the remnant field of the superconducting magnet.

**RESULTS AND ANALYSIS**

We present in figure 1 the dc magnetization from $RuSr_2EuCu_2O_8$ in the low-field and high-field regime. We plot the zero-field-cooled (ZFC) dc susceptibility at 4 mT and the field-cooled dc susceptibility at 6 T. We have previously shown that the decrease in the low-field ZFC susceptibility below ~132 K is a signature of predominantly low-field antiferromagnetic order [16]. The field-cooled susceptibility for low applied fields indicates partial alignment of a weak ferromagnetic component which we estimate to be ~0.05 $\mu_B$/Ru [16]. There is no temperature-dependent hysteresis for an applied magnetic field of 6 T and we have interpreted the magnetization at this field in terms of predominately ferromagnetic order. The appearance of the zero-field Meissner phase is apparent in the insert to figure 1 where we plot the zero-field-cooled ac susceptibility for zero applied field. The decrease in the low-field zero-field-cooled dc susceptibility below ~32 K mirrors that seen in $RuSr_2GdCu_2O_8$ at 45 K and has been attributed to the onset of superconductivity. The appearance of a Meissner phase at a lower temperature in $RuSr_2GdCu_2O_8$ has been attributed to granularity or the appearance of a spontaneous vortex phase [13]. Recent microwave measurements on the $RuSr_2EuCu_2O_8$ sample indicate that the decrease below ~32 K is due to superconductivity and that the Meissner phase is suppressed due to a spontaneous vortex phase [18].

In figure 2 we show the $^{63,65}Cu$ NQR spectra of $RuSr_2EuCu_2O_8$. The spectral positions of $^{63}Cu$ and $^{65}Cu$ are at 29.8 MHz and 27.6 MHz respectively. The spectral linewidths are narrower than in $La_{2-x}Sr_xCuO_4$ [29], comparable with underdoped $YBa_2Cu_3O_{7-\delta}$ [30] but much broader than in stochiometric $YBa_2Cu_4O_8$ [31]. At room temperature we find a linewidth of about 900 kHz for both Cu isotopes. The additional line broadening may be due to structural disorder induced by a rotation of the $RuO_6$ octahedra. We note that the spectral positions of the Cu resonance lines are very close to those observed in underdoped $YBa_2Cu_3O_{7-\delta}$ and the underdoped HTSC, $YBa_2Cu_4O_8$. Since the structure of $RuSr_2EuCu_2O_8$ is very similar to the $YBa_2Cu_3O_{7-\delta}$ family and since there is a linear relation between the hole concentration and Cu NQR resonance position in the same materials family [32] this gives some evidence that this material is underdoped with p=0.127. We note, however, that this analogy is only valid under the assumption that $RuSr_2EuCu_2O_8$ belongs to the $YBa_2Cu_3O_{7-\delta}$ family.

The doping state of the $CuO_2$ planes in $RuSr_2EuCu_2O_8$ can also be estimated from $^{17}O$ NMR measurements. Furthermore, $^{17}O$ NMR measurements enable a direct estimate of the hyperfine and dipolar coupling. Therefore, we now consider the $^{17}O$ NMR spectra in an applied magnetic field of 9 T. As discussed above, the magnetically ordered phase at this field is expected to be predominately ferromagnetic. We show in figure 3 that there are dramatic changes in the $^{17}O$ NMR spectra with decreasing temperature. The high temperature $^{17}O$ NMR spectrum contains a sharp peak as well as broad features extending out to $\pm 2$ MHz. We show below that the sharp peak arises from oxygen in the $CuO_2$ planes. We first note that $^{17}O$ is a spin 5/2 nuclei and hence five transitions are expected. At this field, the quadrupole interaction, in first order perturbation, results in a central $1/2 \leftrightarrow -1/2$ transition as well as four lines arising from the $5/2 \leftrightarrow 3/2$, $3/2 \leftrightarrow 1/2$, $-1/2 \leftrightarrow -3/2$ and $-3/2 \leftrightarrow -5/2$ transitions. For powder samples, the satellite transitions become powder patterns where the singularities are determined by the magnitude and symmetry of the electric field gradient tensor.

The sharp peak in figure 3 arises from oxygen in the $CuO_2$ planes (O(2) sites). The high temperature broad background, extending to $\pm 2$ MHz, is similar to that observed in $YBa_2Cu_3O_{7-d}$ powder samples. It contains the corresponding satellite powder pattern and possibly a contribution from the apical oxygen (O(1) site). The $^{17}$O NMR signal from oxygen in the RuO$_2$ plane (O(3) site) and the apical oxygen are significantly broadened by hyperfine, and as we show later, dipole coupling to the Ru moment. In addition, based on a $^{17}$O NMR study on Sr$_2$RuO$_4$ [33], the $^{17}$O NMR shift from O(3) is expected to be highly anisotropic, which for a powder sample will lead to a very broad central line. Furthermore, a study on ferromagnetic SrRuO$_3$ found that the $^{17}$O NMR linewidth from oxygen in the RuO$_2$ planes is significantly broadened by the Ru moment resulting in a linewidth that is ~150 kHz at 500 K and increases dramatically with decreasing temperature due to hyperfine and dipolar couplings to the Ru moments [34]. We note the that the apical oxygen site is very close to the Ru magnetic moment and hence strongly influenced by the static and dynamic properties of the Ru moment.

It is apparent in figure 3 that the width of the central peak increases dramatically with decreasing temperature but there is no large shift in the position of this peak. The broad signal from the satellite transitions also displays complex behavior with decreasing temperature and there is only a single broad peak at low temperature. We show below that the $^{17}$O NMR data can be interpreted in terms of additional hyperfine and dipolar coupling from the Ru moments to oxygen in the CuO$_2$ layers. To account for the $^{17}$O NMR data we assume that the nuclear Hamiltonian can be written as,

$$^{17}H = H_Z + H_{CS} + H_{SMR} + H_Q + H_{D,Ru-O} + H_{hf,Ru-O}, \qquad (2)$$

where $H_Z$ is the Zeeman term, $H_{CS}$ is the temperature-independent chemical shift term, $H_{SMR}$ is the Shastry-Mila-Rice (SMR) term [35] accounting for hyperfine coupling within the CuO$_2$ plane, $H_Q$ is the nuclear quadrupole term, $H_{D,Ru-O}$ denotes the dipole-dipole interaction between the Ru moment and the $^{17}$O nuclear spin and $H_{hf,Ru-O}$ is an additional term introduced to account for transferred hyperfine coupling from the Ru moments to oxygen in the CuO$_2$ planes. The first four terms are well-known for the HTSC where at 9 T the largest term is the Zeeman term followed by the Quadrupole term which is normally treated in first order perturbation. It has been shown that the HTSC $^{17}$O NMR spectra can be modeled with a $H_{SMR}$ term that involves transferred hyperfine coupling from the spins on the two nearest-neighbor Cu sites to the oxygen site.

For a quantitative estimation of the Ru contribution to the O(2) line position we assume a simple model for $H_{hf,Ru-O}$ of the form $H_{hf,Ru-O} = \hbar ^{17}g \sum_j {}^{17}\hat{\mathbf{I}} \cdot {}^{17}\mathbf{A}_{Ru-O} \cdot \hat{\mathbf{S}}_j$ where, $^{17}g$ is the $^{17}$O magnetogyric ratio, $^{17}\hat{\mathbf{I}}$ is the $^{17}$O nuclear spin, $\mathbf{A}_{Ru-O}$ is the Ru-O hyperfine tensor and $\hat{\mathbf{S}}$ is the Ru spin where it is assumed that the dominant contribution arises from a transferred hyperfine coupling between a particular oxygen site in the CuO$_2$ planes and the two nearest Ru spins. We note that the dipolar interaction $H_{D,Ru-O}$ does not contain any contribution to the line position since the tensor $H_{D,Ru-O}$ is traceless and the measurements were made on a non-oriented powder sample.

From the position of the broad peak at *low temperatures*, we estimate that the isotropic local field is small. This is clear in figure 4 where we plot the $^{17}O$ NMR shift against temperature. We estimate the isotropic $^{17}O$ hyperfine field by comparing the shift at low temperatures with that far above the magnetic ordering temperature to be ~18 mT. However, the low temperature $^{17}O$ NMR linewidth indicates a broadening of ±180 mT. It is possible that the hyperfine coupling tensor is highly anisotropic and the resulting hyperfine field changes sign for Ru moments aligned in different directions with respect to the crystal frame. The large symmetric broadening at low temperatures could then possibly be explained by the ferromagnetic Ru lattice being aligned in the direction of the applied magnetic field but with randomly orientated microcrystallites. The isotropic hyperfine coupling constant can be estimated from $^{17}B_{l,iso} = 2\,^{17}A_{Ru-O,iso} <s>$ where $^{17}B_{l,a}$ is the local hyperfine field, $^{17}A_{Ru-O,iso}$ is the Ru-O isotropic hyperfine coupling constant and $<s>$ is the average Ru spin. As mentioned above, we estimate that $<s>$ is ~1 for an applied magnetic field of 9 T and low temperatures. Hence, the isotropic hyperfine coupling constant, $A_{Ru-O,iso}$ is ~9 mT.

We show below that it is also possible to extract quantitative information concerning the isotropic Cu and Ru contributions to the $^{17}O$ hyperfine shift tensor from the *high temperature* data above the magnetic ordering temperature. In this temperature range we find that the $^{17}O$ NMR shift decreases and then increases with decreasing temperature rather than following a Curie-like behavior as might be expected due to coupling to the Ru moment. The initial decrease in the $^{17}O$ NMR shift is also observed in the underdoped HTSC [36] and provides additional evidence that RuSr$_2$EuCu$_2$O$_8$ is underdoped. The increase in the $^{17}O$ NMR shift for temperature less than ~200 K is not observed in the HTSC and it arises from the Ru moment. Therefore, we model the $^{17}O$ NMR shift as a sum of three terms which accounts for the hyperfine coupling from the Cu spins, the hyperfine coupling from the Ru moments and the temperature independent chemical shift. This can be expressed as [37],

$$^{17}K = \frac{1}{g_{eff,Cu}\,m_B} 2\,^{17}A_{Cu-O,iso}\,c_{Cu}(T) + \frac{1}{g_{eff,Ru}\,m_B} 2\,^{17}A_{Ru-O,iso}\,c_{Ru}(T) + s \,, \qquad (3)$$

where $A_{Cu-O,iso}$ and $A_{Ru-O,iso}$ are the isotropic hyperfine coupling constants and $c_{Cu}(T)$ and $c_{Ru}(T)$ are the spin susceptibilities of Cu and Ru respectively. $g_{eff,Cu}$ and $g_{eff,Ru}$ are the effective g-factors of Cu and Ru (in units of $\mu_B$) and their values are 2 [37] and 3 [16] respectively. $\sigma$ is the temperature independent chemical shift contribution. The susceptibilities, $c_{Cu}(T)$ and $c_{Ru}(T)$, are obtained from the measured molar spin susceptibility, $c^{(molar)}(T)$, which is given by,

$$c^{(molar)}(T) = N_A \left( n_{cu} c_{Cu}(T) + n_{Ru} c_{Ru}(T) \right) + c_{VV,Eu}^{(molar)}, \tag{4}$$

where $N_A$ is Avogadros number, $n_{cu}$ and $n_{Ru}$ are the number of Cu and Ru sites per unit cell and $c_{VV,Eu}^{(molar)}$ is the van-Vleck-contribution due to $Eu^{3+}$ ($^7F_0$). Similar to the high temperature susceptibility study on $RuSr_2EuCu_2O_8$ [38], we use a spin-orbit coupling constant of 303 cm$^{-1}$ as measured in $Eu_2CuO_4$ [39].

For underdoped $YBa_2Cu_3O_{7-d}$, $c_{Cu}(T)$ decreases with decreasing temperature. This decrease has been attributed to the normal-state pseudogap observed in the HTSC [40,41]. There are many expressions for the temperature dependence of spin susceptibility in the HTSC. However, for the temperature range above 140 K we use the simplest phenomenological expression for the the spin susceptibility that has been applied to underdoped $YBa_2Cu_3O_{7-d}$ and is given by [37,42],

$$c_{Cu}(T) = c_0 \left[ 1 - \tanh\left( \frac{T_0}{T} \right) \right], \tag{5}$$

where $T_0$ is a characteristic temperature related to the normal-state pseudogap energy and $c_0$ is a experimental fit parameter.

Using the experimental magnetization data for $c^{(molar)}(T)$ shown in figure 1, which have been extrapolated to 9 T, we were able to fit our experimental $^{17}$O NMR shift data using equations (3), (4) and the phenomenological expression for $c_{Cu}(T)$ given by equation (5). The result of the numerical fit is shown in figure 5a together with the experimental data. We also show in figure 5a the contributions of the three different terms in equation (3). From this analysis, we find an isotropic Cu-O hyperfine coupling constant of 22 T and a Ru-O hyperfine coupling constant of 9 mT. The Cu-O hyperfine coupling constant is comparable with that found in the HTSC [42]. The Ru-O hyperfine coupling constant is consistent with the result obtained from the low temperature shift values discussed above. This means that the contribution of Ru to the hyperfine interaction at the planar oxygen site is more than two thousand times less than the contribution of neighboring planar Cu sites. This result implies that the exchange coupling of the Ru moment to the $CuO_2$ planes is weak.

We note that the fitted characteristic temperature, $T_0$, is 180 K and the fitted chemical shift is 280 ppm. The fitted $T_0$ can be compared to that obtained from $Y_{1-x}Ca_xBa_2Cu_3O_{7-\delta}$ when equation 5 is fitted to the $^{89}$Y NMR shift [40]. We find that $T_0$ increases from 14 K for p=0.173 to 270 K for

p=0.053. Thus, $T_0$ determined from the $^{17}$O NMR shift measurements on RuSr$_2$EuCu$_2$O$_8$ is comparable to very underdoped Y$_{1-x}$Ca$_x$Ba$_2$Cu$_3$O$_{7-\delta}$ with a hole concentration of ~0.07.

It is apparent in the insert of figure 4 and figure 5b that the oxygen linewidth is strongly affected by the Ru moments. For temperatures above the magnetic ordering temperature we find that the temperature dependence of the linewidth (figure 5b) directly follows the susceptibility shown in figure 1. We show below that this behavior is due to Ru-O(2) dipolar coupling rather than any indirect hyperfine coupling from the Ru moment to the O(2) site. Unlike the analysis of the $^{17}$O NMR shift, the dipolar interaction can significantly contribute to the NMR linewidth, especially if an electronic spin is involved. For a Gaussian line, the linewidth δ (FWHM) is given by [43],

$$d = 2\sqrt{2\ln 2 M_2}, \qquad (6)$$

where $M_2$ is the second moment of interaction. For $M_2$ we use the well-known van Vleck expression for dipolar interactions between unlike spins, which is given by [43,44],

$$M_2 = \frac{1}{3}\left(\frac{\mu_0}{4\pi}\right)^2 {}^{17}g_n^2 \, {}^{Ru}g_{el}^2 S(S+1)\hbar^2 \sum_j \frac{(1-3\cos^2\vartheta_j)^2}{d_j^6}, \qquad (7)$$

where $^{17}g_n^2$ denotes the gyromagnetic ratio of $^{17}$O, $^{Ru}g_{el}^2$ is the gyromagmetic ratio of the Ru spin, S is the Ru spin, $\vartheta_j$ measures the angle between the field direction and the direction of the Ru-O vector and $d_j$ is the distance between the oxygen site under consideration and a particular Ru site. The sum in equation (7) comprises all Ru sites.

We now introduce a temperature dependence of $M_2$ by the following substitution in equation (7),

$$^{Ru}g_{el}\sqrt{S(S+1)} \rightarrow \frac{g_{eff,Ru}\mu_B B_0}{3k_B(T-\theta)}\frac{g_{eff,Ru}\mu_B}{\hbar}. \qquad (8)$$

The physical meaning of this substitution is that a particular $^{17}$O nuclear spin experiences a temperature and field dependent average Ru moment from each Ru site which is described by a Curie-Weiss-like term. This model is valid for exchange-coupled localized Ru moments and temperatures above the magnetic ordering transition temperature, θ. Using the parameters obtained by fitting the high temperature magnetization data (θ=122K, $g_{eff,Ru}$=3 [16]), and calculating the lattice sum in equation (7) numerically using the structure data of RuSr$_2$EuCu$_2$O$_8$ [5], we are able to calculate the temperature dependence of the dipolar linewidth contribution at the different oxygen sites when $B_0$=9 T. For the planar oxygen site in the CuO$_2$ plane, we find a temperature dependence shown by the solid line in figure 5b. Although the theoretical curve is slightly above the experimental data (about 10%), over the whole temperature range the experimental data are very well described by this model. The small deviation is likely to be due to the changing effective Ru moment with decreasing temperature. It is known from magnetization studies that the effective Ru moment decreases at lower temperatures and a Currie-Weiss approximation is only strictly valid above 200 K [16]. For the apical oxygen site O(1) and the planar oxygen site O(3) in the RuO$_2$ lattice, we find oxygen linewidths at room temperature which are almost an order of magnitude (7.5 for O(1) and 10 for O(3)) larger than that for the O(2) site in the CuO$_2$ planes. Therefore, even before considering the hyperfine interactions, it is clear that dipole interactions lead to significant broadening of the O(1) and O(3) NMR lines.

Unlike the $^{17}$O NMR shift, but similar to the $^{17}$O NMR linewidth, we find that $1/^{17}T_1T$ from O(2) is strongly affected by coupling to the Ru moment. This is apparent in figure 4 (right axis) were it can be seen that $1/^{17}T_1T$ increases with decreasing temperature. This indicates that there is additional relaxation via hyperfine and dipolar coupling from the Ru moment because in underdoped YBa$_2$Cu$_3$O$_{7-\delta}$, $1/^{17}T_1T \propto c(T)$ and hence $1/^{17}T_1T$ is found to decrease with decreasing temperature [36]. The Curie-like increase in $1/^{17}T_1T$, observed in RuSr$_2$EuCu$_2$O$_8$, can be understood by noting that there is hyperfine coupling from the two nearest-neighbor Cu spins to oxygen in the CuO$_2$ planes. Therefore, $1/^{17}T_1T$ probes the imaginary part of the dynamical spin susceptibility near $\mathbf{q}=(0,0)$ in the CuO$_2$ planes. From the assumed symmetry of the hyperfine coupling of oxygen in the CuO$_2$ planes, it is also possible that $1/^{17}T_1T$ from coupling to the Ru moments, predominately probes the $\mathbf{q}=(0,0)$ spin dynamics of the Ru lattice. This will lead to the additional Curie-like term. It is equally possible that the $^{17}$O spin-lattice relaxation is governed by an additional term stemming from a dipolar relaxation mechanism between the Ru moment and the $^{17}$O nuclear spin.

We note that we were not able to extend our quantitative models for the temperature dependence of the $^{17}$O NMR shift and the $^{17}$O NMR linewidth presented above and in figure 5 to temperatures below the magnetic ordering temperature. This implies that the hyperfine and dipolar couplings between Ru and $^{17}$O in the CuO$_2$ plane change in the magnetically ordered phase. We have shown in a previous magnetization study that the effective Ru moment changes from 3 $\mu_B$ at high temperatures to 1 $\mu_B$ at low temperatures [16]. This directly affects the $^{17}$O NMR shift as can be seen from the second term in equation (3). However, we find that there is no significant change in the isotropic Ru-O hyperfine coupling constant obtained at temperatures above the magnetic ordering transition temperature using an effective Ru moment of 3 $\mu_B$ and at very low temperatures using an effective Ru moment of 1 $\mu_B$. The gradual change of the effective Ru magnetic moment also affects the $^{17}$O NMR linewidth as can be seen from equation (8). The modeling of the temperature dependence of the $^{17}$O NMR linewidth is further complicated when the linewidth of the central peak equals the quadrupolar frequency since an admixture of satellite transitions will then occur.

**CONCLUSION**

In conclusion, we have performed $^{17}$O NMR measurements on RuSr$_2$EuCu$_2$O$_8$, which displays both superconductivity and magnetic order. The $^{17}$O NMR shift above the magnetic ordering temperature is only weakly affected by coupling to the RuO$_2$ layers and the NMR shift above 200 K increases with increasing temperature. By fitting the $^{17}$O NMR shift above the magnetic ordering transition temperature, we find that the hole concentration on the CuO$_2$ planes in RuSr$_2$EuCu$_2$O$_8$ is similar to a very underdoped HTSC. Furthermore, we find a small isotropic Ru-O(2) hyperfine coupling constant of 9 mT above and below the magnetic ordering temperature. Both the $^{17}$O NMR linewidth and $1/^{17}T_1T$ are strongly affected by the Ru moments in the RuO$_2$ layers for temperatures up to, and above, 300 K. We find that the temperature dependence of the $^{17}$O NMR linewidth can be attributed to dipolar coupling from the moments in the RuO$_2$ layers. It is possible that dipolar coupling and hyperfine coupling cause the increase in $1/^{17}T_1T$ with decreasing temperature for temperatures above the magnetic ordering transition temperature.

**Acknowledgements**

We acknowledge funding support from the New Zealand Marsden Fund (GVMW) and the Alexander von Humboldt Foundation (GVMW). We thank M. Mehring, J. Haase, E. Pavarini and Y. Tokunaga for stimulating discussions and B. Walker, J. L. Tallon and C. Bernhard for providing and processing the samples.

**FIGURES**

**Figure 1:** Plot of the dc magnetic susceptibility against temperature for applied dc magnetic fields of 4 mT and 6 T. Insert: Plot of the zero-field ac magnetic susceptibility against temperature for an ac excitation field of 5 µT and a frequency of 1 kHz.

**Figure 2**: Plot of the Cu NQR spectra at room temperature. Two peaks are visible which can be attributed to $^{63}$Cu and $^{65}$Cu sites. The dashed line is a fit using two Gaussian lines.

**Figure 3:** Plot of the $^{17}$O NMR spectra against frequency for the temperatures indicated and an applied magnetic field of 9 T. The frequency is referenced to $H_2O$ (52.453 MHz).

**Figure 4:** Plot of the $^{17}$O NMR shift (filled circles, left axis) and $(^{17}T_1T)^{-1}$ (open circles, right axis) against temperature. The $^{17}$O NMR shift was referenced to $H_2O$. Insert: Plot of the $^{17}$O NMR linewidth against temperature.

**Figure 5:** Plot of the $^{17}$O NMR shift (a) and linewidth (b) data. The lines are fits to the model described in the text. The solid line is the total shift fitted using equation (3) and the experimental susceptibility data shown in figure 1. Also shown are the contributions from Cu (dashed line), Ru (dotted line) and the chemical shift (dash-dot-dash line).

Figure 1
Krämer and Williams

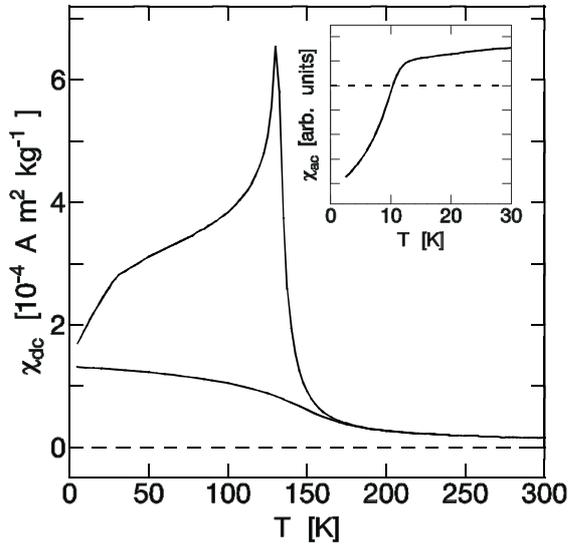

Figure 2
Krämer and Williams

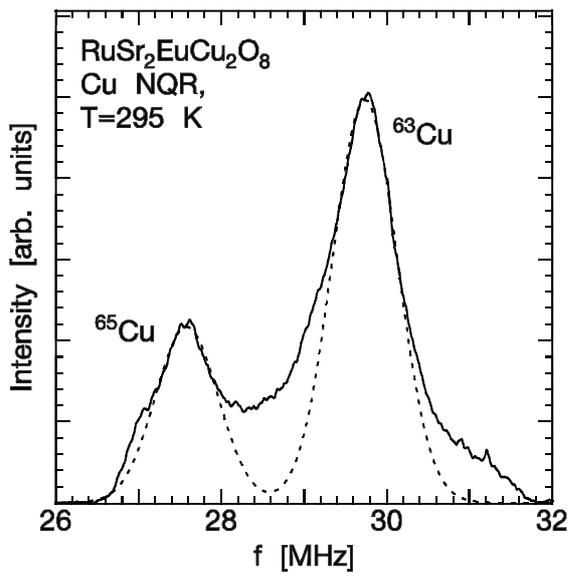

Figure 3
Krämer and Williams

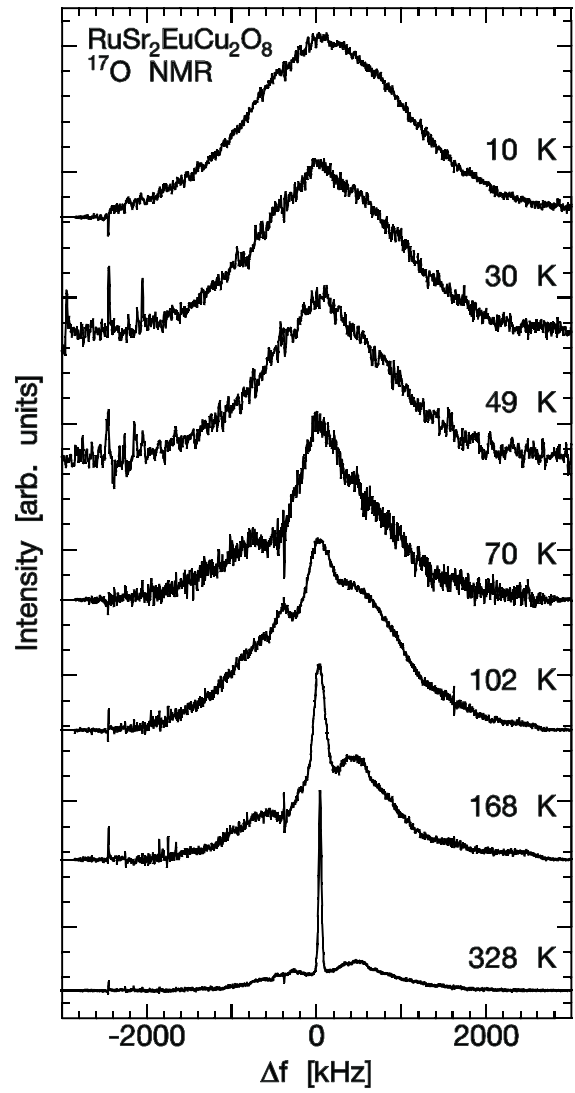

Figure 4
Krämer and Williams

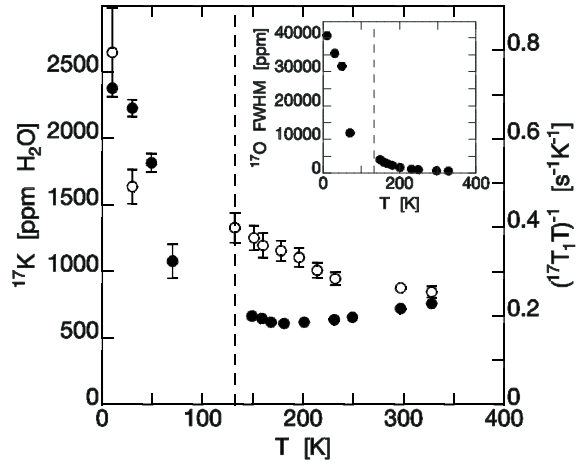

Figure 5
Krämer and Williams

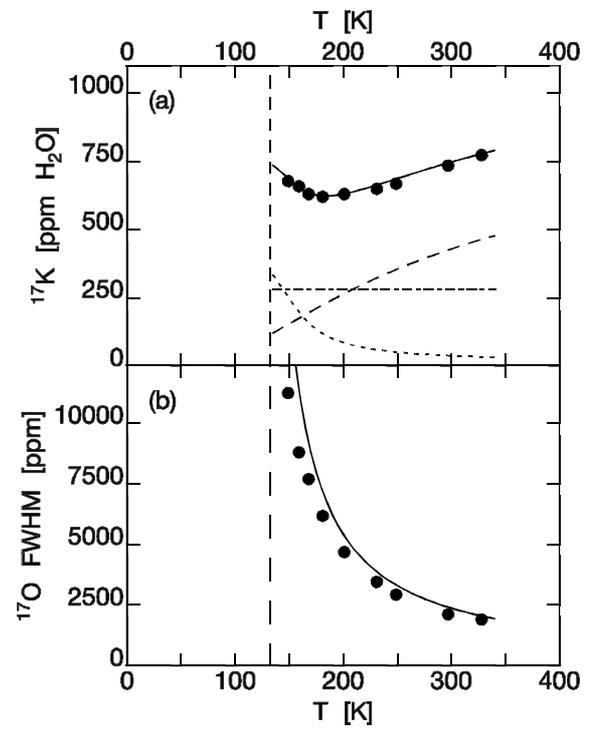